\documentclass[nofootinbib,amsmath,twocolumn,notitlepage,preprintnumbers]{revtex4-1}
\usepackage{multirow}
\usepackage{amssymb, esvect, amsmath, graphicx, latexsym, amsthm, slashed, eso-pic}
\usepackage{xcolor}
\usepackage{hyperref}
\usepackage{amsmath}

\newcommand{\beq}{\begin{equation}} \newcommand{\eeq}{\end{equation}}
\newcommand{\bea}{\begin{eqnarray}} \newcommand{\eea}{\end{eqnarray}}

\def\lsim{\mathrel{\raise.3ex\hbox{$<$\kern-.75em\lower1ex\hbox{$\sim$}}}}
\def\gsim{\mathrel{\raise.3ex\hbox{$>$\kern-.75em\lower1ex\hbox{$\sim$}}}}

\newcommand{\be}{\begin{eqnarray}}
\newcommand{\ee}{\end{eqnarray}}

\newcommand{\benum}{\begin{enumerate}}
\newcommand{\eenum}{\end{enumerate}}
\newcommand{\bi}{\begin{itemize}}
\newcommand{\ei}{\end{itemize}}

\begin{document}

\preprint{FERMILAB-PUB-22-251-T}

\title{The Sensitivity of Future Gamma-Ray Telescopes to Primordial Black Holes}

\author{Celeste Keith$^{a,b}$}
\thanks{ORCID: https://orcid.org/0000-0002-3004-0930}

\author{Dan Hooper$^{a,b,c}$}
\thanks{ORCID: http://orcid.org/0000-0001-8837-4127}

\author{Tim Linden$^{d}$}
\thanks{ORCID: http://orcid.org/0000-0001-9888-0971}

\author{Rayne Liu$^{a, b}$}
\thanks{ORCID: http://orcid.org/0000-0003-3274-8964}

\affiliation{$^a$University of Chicago, Kavli Institute for Cosmological Physics, Chicago IL, USA}
\affiliation{$^b$University of Chicago, Department of Astronomy and Astrophysics, Chicago IL, USA}
\affiliation{$^c$Fermi National Accelerator Laboratory, Theoretical Astrophysics Group, Batavia, IL, USA}
\affiliation{$^d$Stockholm University and The Oskar Klein Centre for Cosmoparticle Physics, AlbaNova, 10691 Stockholm, Sweden}

\date{\today}

\begin{abstract}

The strongest existing constraints on primordial black holes with masses in the range of \mbox{$m_{\rm BH} \sim 10^{15}-10^{17} \, {\rm g}$} have been derived from measurements of the local cosmic-ray electron-positron flux by Voyager 1, and MeV-scale gamma-ray observations of the Inner Galaxy by COMPTEL and INTEGRAL. In this paper, we evaluate the sensitivity of future MeV-scale gamma-ray telescopes such as e-ASTROGAM or AMEGO to Hawking radiation. We show that such an instrument would be able to provide the strongest constraints on black holes in the mass range of $m_{\rm BH} \sim (0.6-20) \times 10^{16} \, {\rm g}$, typically exceeding current constraints by approximately two orders of magnitude. In scenarios in which the observed 511 keV excess is the result of Hawking radiation, we find that e-ASTROGAM or AMEGO would not only be able to detect the Hawking radiation from the Inner Galaxy, but could precisely measure the abundance and mass distribution of the black holes responsible for this signal.

\end{abstract}

\maketitle

\section{Introduction}

If the early universe contained large curvature perturbations, a sizable population of primordial black holes may have formed. While various constraints have been placed on the characteristics of any such population of black holes (for a review, see Ref.~\cite{Carr:2020gox}), significant abundances of such objects may still be present in our universe today. 

It has long been appreciated that black holes radiate an approximately thermal spectrum of particles, known as Hawking radiation~\cite{Hawking:1974sw,Gibbons:1977mu}. Searches for Hawking radiation in the form of gamma rays and positrons have been used to derive powerful constraints on primordial black holes~\cite{Keith:2021guq,Coogan:2020tuf,Laha:2020ivk,Boudaud:2018hqb,Dasgupta:2019cae}. More specifically, local measurements of the cosmic-ray electron (plus positron) flux by the Voyager 1 satellite currently provide the strongest constraint on black holes lighter than $m_{\rm BH }\sim (1-2) \times 10^{16} \, {\rm g}$~\cite{Boudaud:2018hqb}, while MeV-scale gamma-ray observations of the Inner Galaxy by COMPTEL and INTEGRAL provide the leading constraints in the mass range of $m_{\rm BH}\sim 10^{16}-10^{17} \, {\rm g}$~\cite{Keith:2021guq,Coogan:2020tuf,Laha:2020ivk,Coogan:2020tuf}.\footnote{While GeV-scale telescopes such as EGRET~\cite{Lehoucq:2009ge} and Fermi~\cite{Fermi-LAT:2018pfs} can be used to search for the Hawking radiation from lower mass black holes, the constraints provided by such instruments are less stringent than those derived from Voyager 1's measurements of the local electron-positron flux~\cite{Boudaud:2018hqb}.}

While the data provided by the COMPTEL~\cite{Strong:1998ck}  and INTEGRAL~\cite{Bouchet:2011fn} satellites have made it possible to derive interesting bounds on the abundance of primordial black holes, the sensitivity of such instruments is limited. Fortunately, a new generation of satellite-based MeV-scale gamma-ray telescopes have been proposed, including the designs currently known as AMEGO (All-sky Medium Energy Gamma-ray Observatory)~\cite{McEnery:2019tcm} and e-ASTROGAM (``enhanced ASTROGAM'')~\cite{DeAngelis:2016slk}. Such instruments would be capable of detecting photons through both pair conversion (as Fermi, for example, does) and Compton scattering, enabling them to have much greater sensitivity to photons in the 1-100 MeV range. While COMPTEL and INTEGRAL are able to detect MeV-scale photons, the projected sensitivity of AMEGO and e-ASTROGAM to such gamma rays exceeds that of these earlier instruments by roughly two orders of magnitude.

In this paper, we consider the sensitivity of next-generation MeV-scale gamma-ray telescopes to the Hawking radiation from a population of primordial black holes (for earlier related work, see Ref.~\cite{Coogan:2020tuf}). To this end, we have calculated the energy spectrum and angular distribution of the Hawking radiation from a $40^{\circ}\times 40^{\circ}$ region around the Galactic Center, including contributions from inflight electron-positron annihilation and final state radiation. We then performed an analysis of simulated e-ASTROGAM data, utilizing spatial templates, allowing us to fully exploit the morphological and spectral information provided by such an instrument (we expect to obtain similar sensitivity for an instrument such as AMEGO). Through this analysis, we have been able to derive projected constraints on the abundance of primordial black holes in the mass range of \mbox{$m_{\rm BH} \sim (0.3-30)\times 10^{16} \, {\rm g}$}, and for a wide range of halo profiles. For black holes in the mass range of \mbox{$m_{\rm BH} \sim 10^{16}-10^{17} \, {\rm g}$}, we find that such a telescope would be able to improve upon current constraints by approximately a factor of $\sim$\, 100, potentially excluding scenarios in which the black holes make up more than \mbox{$f_{\rm BH} \sim 10^{-4}-10^{-6}$} of the total dark matter density. We also consider scenarios in which primordial black holes are responsible for the excess of 511 keV photons observed from the Inner Galaxy~\cite{Keith:2021guq}, as reported by the INTEGRAL Collaboration~\cite{Churazov:2004as, Jean:2005af,Prantzos:2005pz}. We find that in such a scenario, an instrument such as AMEGO or e-ASTROGAM would not only be able to detect the gamma rays radiated from the black holes, but would be able to quite precisely measure the abundance and mass distribution of the responsible black hole population.

\section{Hawking Radiation From Primordial Black Holes}

The temperature and Schwarzschild radius of a black hole are related to its mass as follows:\footnote{Throughout this study, we will focus our attention on the case of Schwarzschild black holes.}
\begin{align} 
\label{masstemp}
T_{\rm BH} &= \frac{M_{\text{Pl}}^2}{8 \pi m_{\text{BH}}} \approx 1.05 \, {\rm MeV}  \times \bigg(\frac{10^{16} \, {\rm g}}{m_{\rm BH}}\bigg) \\
r_s &=\frac{2m_{\rm BH}}{M^2_{\rm Pl}} \simeq 1.5 \times 10^{-12} \, {\rm cm} \times \bigg(\frac{m_{\rm BH}}{10^{16} \, {\rm g}}\bigg),\nonumber
\end{align}
where $M_{\text{Pl}} \approx 1.22 \times 10^{19} \, {\rm GeV}$ is the Planck mass.

Due to the gravitational nature of Hawking evaporation, the radiation from a black hole includes all species of particles that are lighter than or comparable in mass to the black hole's temperature, leading to the following rate of mass loss:
\be \label{dmdt}
\!\begin{aligned}
 \frac{dm_{\textrm{BH}}}{dt} &= -\frac{\mathcal{G}g_{*,H}(m_{\textrm{BH}})M_{\textrm{Pl}}^2}{30720\pi m_{\textrm{BH}}^2} \\
& \approx -8.2 \times 10^{-7} \,\textrm{g/s}  \times \left(\frac{g_{*,H}}{10.92} \right) \left(\frac{10^{16}\textrm{g}}{m_{\textrm{BH}}} \right)^2, \\
\end{aligned} 
\ee
where $\mathcal{G} \approx$ 3.8 is the greybody factor, and $g_{*,H}$ counts the number of spin-weighted degrees-of-freedom that are lighter than the black hole's temperature. The quantity $g_{*,H}$ receives a contribution of 6 from the three Standard Model neutrinos and antineutrinos, 4 from electrons and positrons, 0.82 from photons, and 0.1 from gravitons~\cite{MacGibbon:1990zk,MacGibbon:1991tj}. Integrating this expression, we find that a black hole with an initial mass of $m_{\rm BH} \sim 4 \times 10^{14} \, {\rm g}$ will evaporate in a length of time equal to the age of the universe. 

The spectrum of Hawking radiation from an individual black hole can be written as follows~\cite{Page:1976df}:
\be 
\label{page}
\frac{dN^{\rm dir}}{dE}(m_{\text{BH}}, E) = \frac{1}{2\pi^{2}}\frac{E^{2}\sigma(m_{\text{BH}},E)}{e^{E/T}\pm1},
\ee
where for fermions (bosons), the sign in the denominator is positive (negative). The absorption cross section, $\sigma$, also depends on the spin of the particles being radiated. In the $E \gg T$ limit, the absorption cross section approaches $\sigma \simeq 27 \pi m^2_{\rm BH}/M^4_{\rm Pl}$, regardless of the particle species. At lower energies, $\sigma$ is a function of energy, and depends on the particle species under consideration. Throughout our calculations, we implement the full spectra as presented in Ref.~\cite{Page:1976df}.


Black holes can produce gamma rays not only as the direct products of Hawking evaporation (as described by Eq.~\ref{page}), but also through the inflight annihilation of positrons, and as final state radiation. The spectrum of gamma rays from the inflight annihilation of positrons is given by~\cite{Beacom:2004pe}  
\begin{widetext}
\begin{align}
\label{eq:IA}
\frac{dN^{\rm IA}_{\gamma}}{dE_{\gamma}} &= \frac{\pi \alpha^2 n_H}{m_e}  \int^{\infty}_{m_e} dE_{e^+} \frac{dN_{e^+}}{dE_{e^+}} \int^{E_{e^+}}_{E_{\rm min}} \frac{dE}{dE/dx} \, \frac{P_{E_{e^+} \rightarrow E} }{(E^2-m_e^2)} \\[3pt] 
&\times  \bigg(-2-\frac{(E+m_e)(m_e^2(E+m_e)+E_{\gamma}^2(E+3m_e)-E_{\gamma}(E+m_e)(E+3m_e))}{E_{\gamma}^2(E-E_{\gamma}+m_e)^2} \nonumber
\bigg), \nonumber 
\end{align}
\end{widetext}
where $\alpha \approx 1/137.037$ is the fine structure constant, $n_H$ is the number density of neutral hydrogen atoms, and $dN_{e^+}/dE_{e^+}$ is the spectrum of positrons radiated from the black hole, as described by Eq.~\ref{page}. The energy loss rate for a positron due to ionization in the presence of neutral hydrogen, $dE/dx$, is given by the standard Bethe-Bloch formula. Note that since $dE/dx$ is proportional to $n_H$, the flux of gamma rays from inflight annihilation is not sensitive to the density of gas.

The probability that a positron with an initial energy of $E_{e^+}$ will survive until its energy has been reduced to $E$ is given by
\begin{align}
P_{E_{e^+}\rightarrow \, E} =\exp\bigg(-n_H \int^{E_{e^+}}_E \sigma_{\rm ann}(E') \,\frac{dE'}{|dE'/dx|}  \bigg),
\end{align}
where $\sigma_{\rm ann}$ is the cross section for a positron to annihilate with an electron at rest. For positrons less energetic than a few MeV, $P_{E_{e^+} \rightarrow \, m_e}$ always falls in the range between 0.95 and 1.0, reflecting the fact that only a few percent of the positrons annihilate before becoming non-relativistic.

\begin{figure*}
\includegraphics[width=3.25in,angle=0]{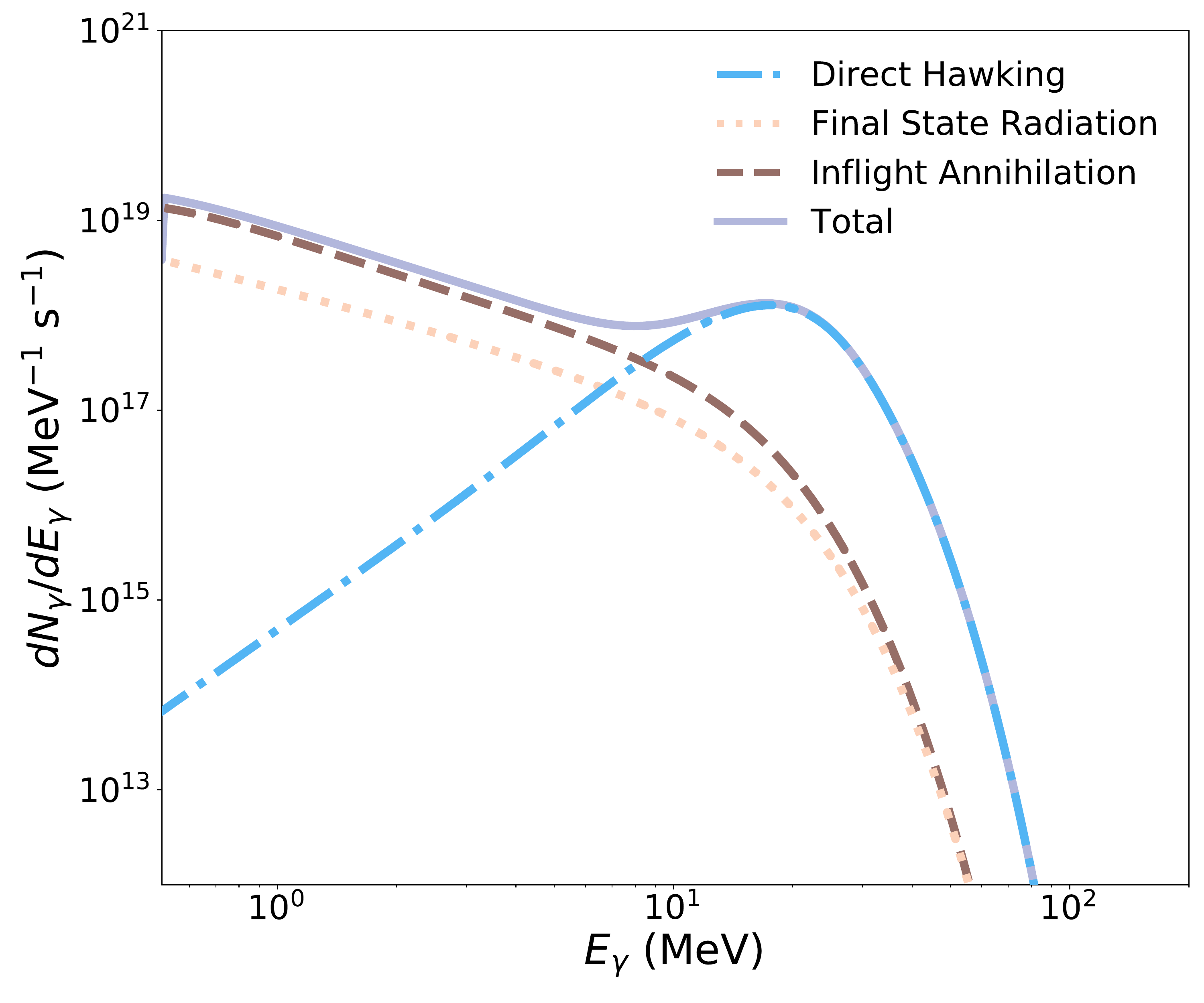}
\includegraphics[width=3.25in,angle=0]{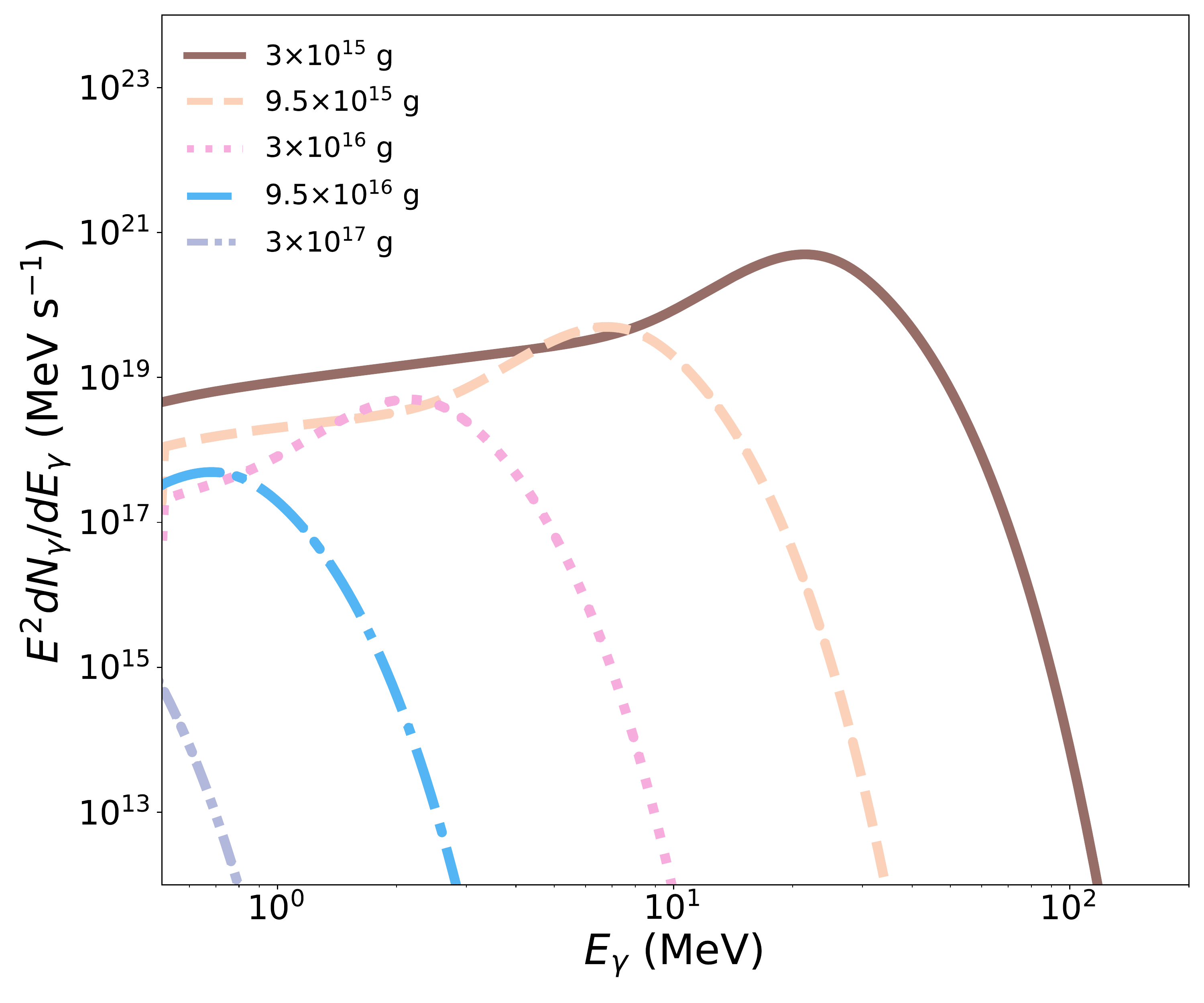}
\caption{Left: The gamma-ray spectrum from a black hole with a mass of $m_{\rm BH} = 3 \times 10^{15} \, {\rm g}$, including the contributions from direct Hawking radiation, final state radiation, and the inflight annihilation of positrons. Right: The total gamma-ray spectrum from black holes for several choices of $m_{\rm BH}$.}
\label{all_three}
\end{figure*}

Lastly, we also include in our calculations the final state radiation from any electrons and positrons that are produced through the process of Hawking evaporation. This leads to the following contribution to the gamma-ray spectrum:
\begin{widetext}
\begin{align}
\frac{dN^{\rm FSR}_{\gamma}}{dE_{\gamma}} &=\frac{\alpha}{2\pi} \int dE_{e} \frac{dN_e}{dE_e} \,  \bigg( \frac{2}{E_{\gamma}} +\frac{E_{\gamma}}{E^2_e} - \frac{2}{E_e} \bigg) \, \bigg[\ln\bigg(\frac{2E_e(E_e-E_{\gamma})}{m^2_e}\bigg)-1\bigg], 
\end{align}
\end{widetext}
where $dN_{e}/dE_{e}$ is the spectrum of electrons and positrons radiated from the black hole.

In Fig.~\ref{all_three}, we show the spectrum of the gamma-ray emission from an individual black hole for several choices of $m_{\rm BH}$, and including contributions from direct Hawking radiation, final state radiation, and inflight annihilation. At the highest energies, direct Hawking radiation dominates this emission. In contrast, inflight annihilation provides the largest contribution at lower energies.  
 
Putting these contributions together, we are now in a position to calculate the total flux of gamma-rays from a population of primordial black holes. Averaged over a solid angle, $\Delta \Omega$, this flux is given by:
\begin{align} 
\label{jfactor}
F_{\gamma}(\Delta \Omega) &= \frac{dN^{\rm tot}_{\gamma}}{dE_{\gamma}}\frac{1}{4\pi} \int_{\Delta \Omega} \int_{los} n_{\textrm{BH}}(l, \Omega) \,dl \, d\Omega, \\
&= \frac{dN^{\rm tot}_{\gamma}}{dE_{\gamma}}\frac{f_{\rm BH}}{4\pi m_{\rm BH}} \int_{\Delta \Omega} \int_{los} \rho_{\textrm{DM}}(l, \Omega) \,dl \, d\Omega, \nonumber
\end{align}
where 
\begin{align}
\frac{dN^{\rm tot}_{\gamma}}{dE_{\gamma}} = \frac{dN^{\rm dir}_{\gamma}}{dE_{\gamma}} +\frac{dN^{\rm IA}_{\gamma}}{dE_{\gamma}} +\frac{dN^{\rm FSR}_{\gamma}}{dE_{\gamma}},
\end{align}
$n_{\rm BH}$ is the number density of black holes, $f_{\rm BH}$ is the fraction of the dark matter that consists of black holes, and the integrals are performed over the line-of-sight and the solid angle observed. For the spatial distribution of primordial black holes in the Milky Way, we adopt a generalized Navarro-Frenk-White (NFW) halo profile \cite{Navarro:1995iw,Navarro:1996gj}:
\begin{align}
n_{\rm BH} =\frac{n_0}{(r/R_s)^{\gamma} \, [1+(r/R_s)]^{3-\gamma}},
\end{align}
where $r$ is the distance from the Galactic Center. In our calculations, we adopt a scale radius of $R_s=20$ kpc and have normalized $n_0$ such that the local density of black holes (at $r=8.25$ kpc) is $n_{\rm BH} = 0.4 \, {\rm GeV/cm}^3 \times f_{\rm BH}/m_{\rm BH}$~\cite{Read:2014qva}. We take the inner slope of this profile, $\gamma$, to be a free parameter.

\begin{figure*}
\includegraphics[width=7.0in,angle=0]{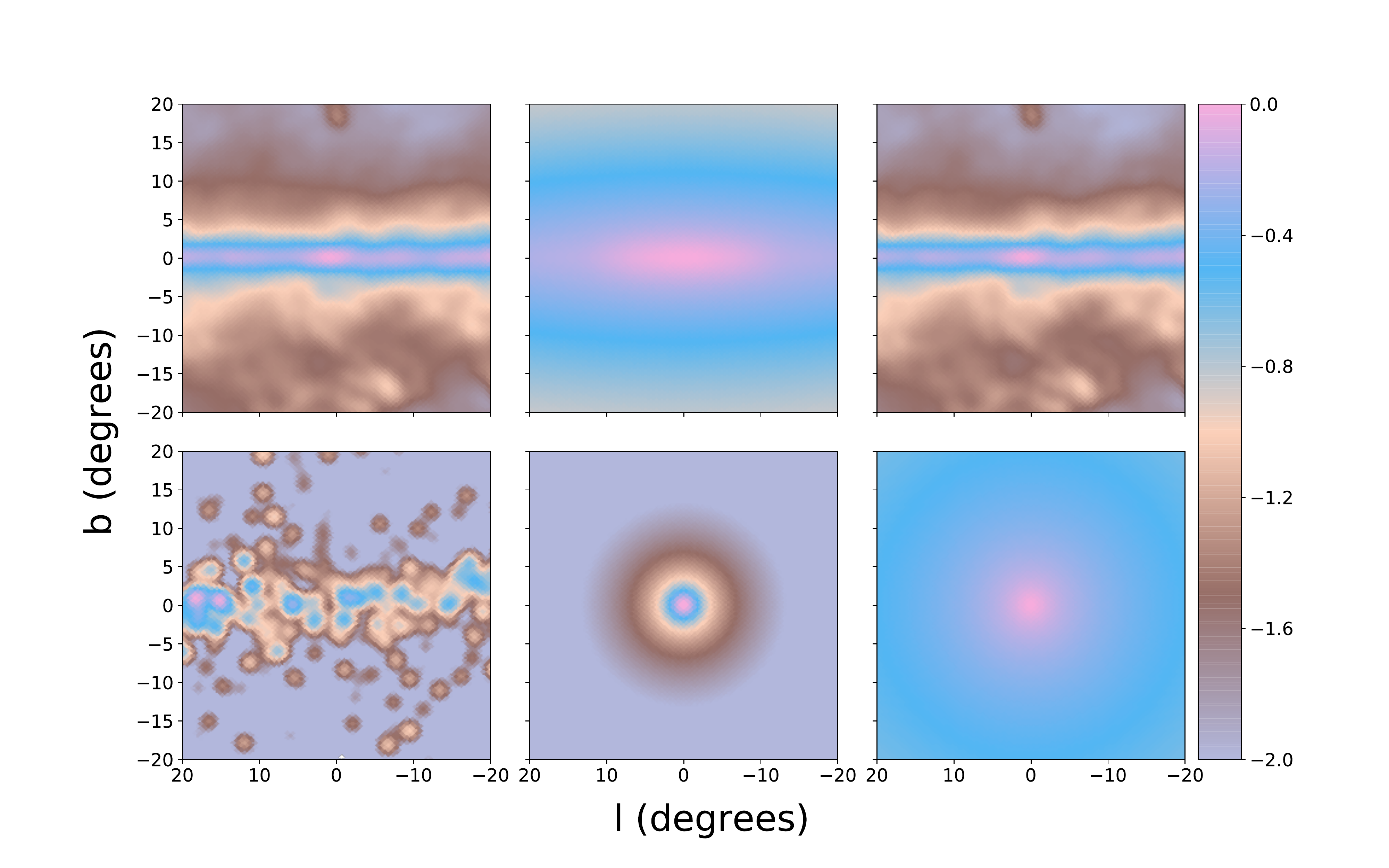}
\caption{The spatial templates used in our analysis evaluated at 10 MeV after convolving with the point spread function of e-ASTROGAM. In the upper row, the templates correspond to the emission from pion production (left), inverse Compton scattering (center), and bremsstrahlung (right), as generated using the publicly available code GALPROP~\cite{2011CoPhC.182.1156V,2005ApJ...622..759G}. In the lower row, the templates correspond to the gamma-ray point sources contained within the Fermi 4FGL-DR2 catalog (left), the emission associated with the Galactic Center gamma-ray excess (center), and the emission from primordial black holes (with $\gamma=1.4$, $m_{\rm BH}=2 \times 10^{16} \, {\rm g}$ and $f_{\rm BH}=10^{-4}$).  The scale used is logarithmic, and the brightest point in each frame is normalized to unity.}
\label{templates}
\end{figure*}

\section{Data Simulation and Template Analysis} 

In order to project the sensitivity of a future MeV-scale gamma-ray telescope to the Hawking radiation from a population of primordial black holes in the Inner Galaxy, we have created a series of simulated data sets based on the proposed design of e-ASTROGAM, and analyzed this simulated data utilizing a number of spatial templates. Such template-based analyses are extremely powerful in that they allow us to simultaneously exploit both spectral and morphological distinctions between the signal being searched for and the various astrophysical backgrounds that are present.

\begin{figure}
\includegraphics[width=3.25in,angle=0]{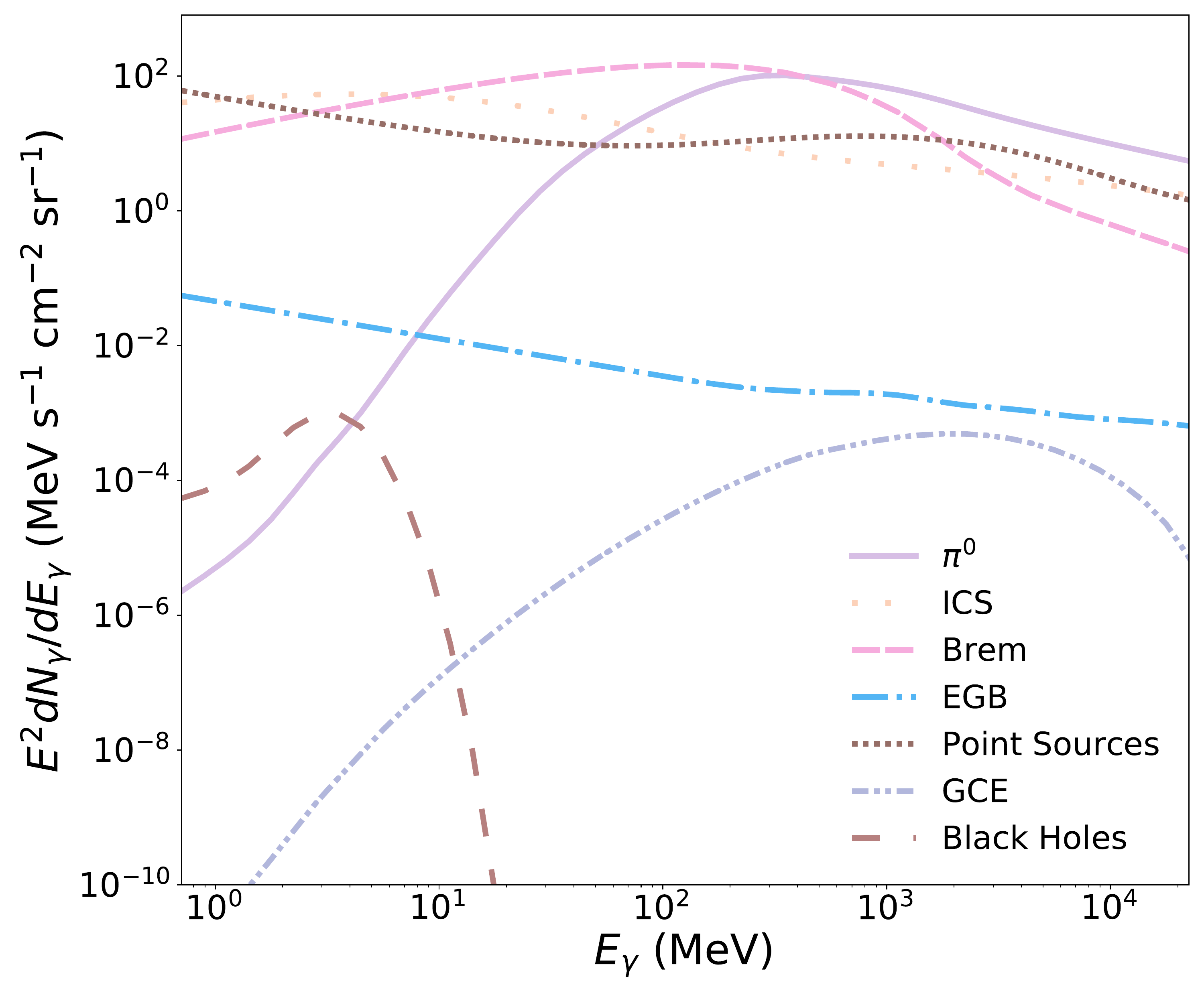}
\caption{The gamma-ray spectra of the various components of our background model, and from primordial black holes (for the case of $m_{\rm BH}=2\times 10^{16} \, {\rm g}$, $f_{\rm BH} =4 \times 10^{-4}$ and $\gamma=1$). Each curve is averaged over the $40^{\circ}\times 40^{\circ}$ region-of-interest.}
\label{templates2}
\end{figure}

Our analysis includes spatial templates associated with the processes of pion production, inverse Compton scattering, and Bremsstrahlung, each of which we generated using the publicly available code GALPROP~\cite{2011CoPhC.182.1156V,2005ApJ...622..759G}.\footnote{In utilizing GALPROP, we have adopted the default parameters from GALPROP WebRun, \url{https://galprop.stanford.edu/webrun.php}, which have been selected to reproduce a variety of cosmic-ray and gamma-ray data.} In addition to these three templates associated with Galactic diffuse emission mechanisms, we have also included templates designed to account for known gamma-ray point sources (using the best-fit spectra and source locations, as reported in the Fermi 4FGL-DR2 catalog~\cite{Fermi-LAT:2019yla}), and for the (isotropic) extragalactic gamma-ray background~\cite{Fermi-LAT:2014ryh}. For each of these two latter templates, we have extrapolated in energy from the range measured by Fermi. We have also included a template intended to account for the emission known as the Galactic Center gamma-ray excess~\cite{Hooper:2010mq,Goodenough:2009gk,Hooper:2011ti,Daylan:2014rsa,Fermi-LAT:2017opo}, which we have modeled as a population of 40 GeV dark matter particles annihilating (to $b\bar{b}$) with a cross section of $\langle \sigma v\rangle = 2.2 \times 10^{-26} \, {\rm cm}^3/{\rm s}$, and that is distributed according to a halo profile with an inner slope of $\gamma=1.2$~\cite{Cholis:2021rpp,Daylan:2014rsa,Fermi-LAT:2017opo}. In this paper, we take no stance on the origin of this excess, which can be treated without loss of generality as arising from the annihilation of particle dark matter, a large population of millisecond pulsars, or from some other unknown process or mechanism. The morphology of these templates, as evaluated at 10 MeV, is shown in Fig.~\ref{templates}. The scale used is logarithmic (base 10), and the brightest point in each frame is normalized to unity. For example, the brightest point in each frame is pink, while a point that is fainter by two orders of magnitude would appear purple. The gamm-ray spectrum associated with each of these templates is shown in Fig.~\ref{templates2}.

In our analysis, we adopt as our region of interest a $40^{\circ} \times 40^{\circ}$ square centered on the Galactic Center. We divide this region into 0.2098 square degree HEALPix bins, corresponding to $N_{\rm side}=128$. We also divide the spectrum into 10 energy bins per decade, from $E_{\gamma}=m_e$ up to 1 GeV.

\begin{figure}
\includegraphics[width=3.25in,angle=0]{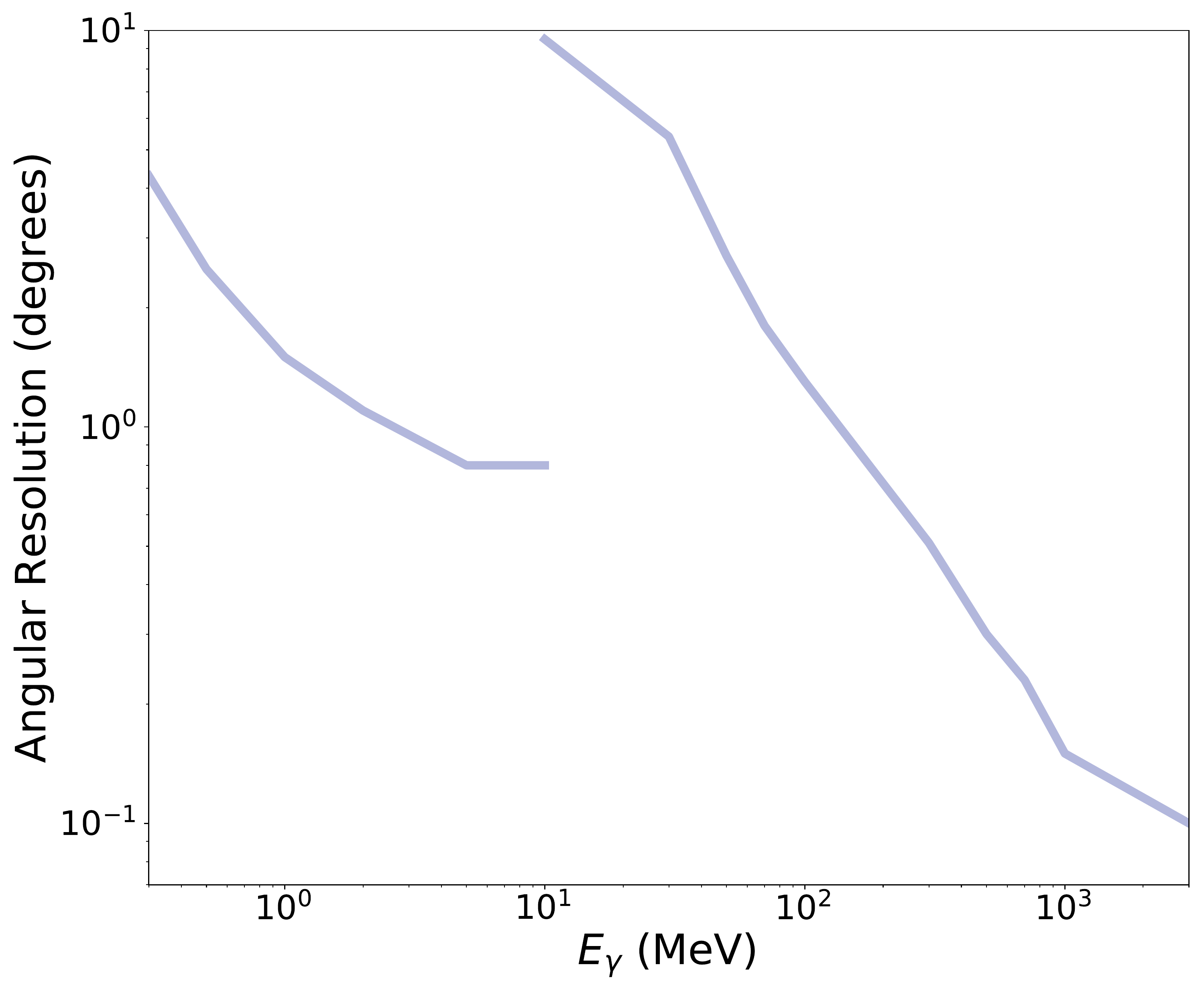} \\
\includegraphics[width=3.25in,angle=0]{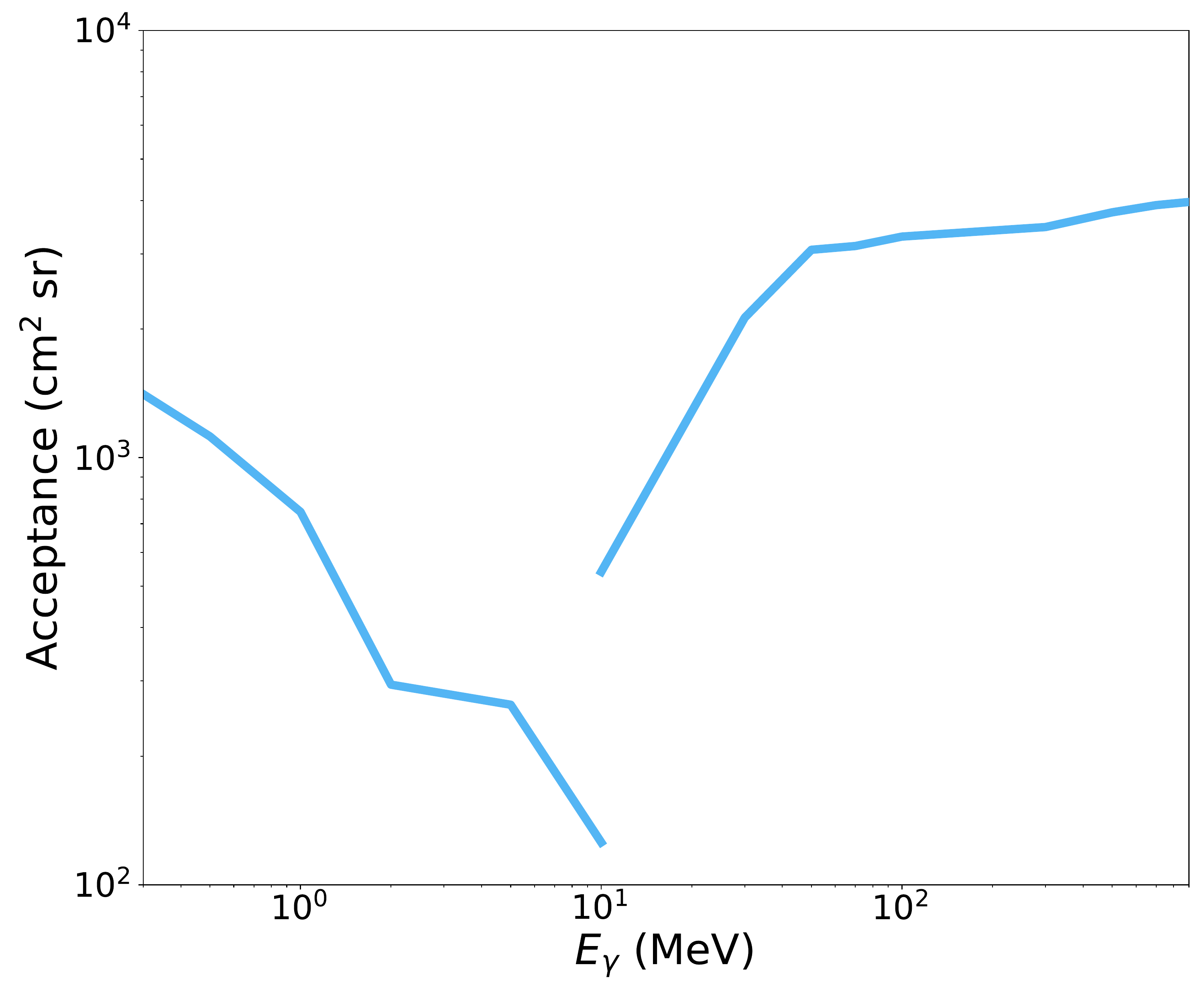}
\caption{The 68\% containment radius (top) and acceptance (bottom) of e-ASTROGAM as a function of gamma-ray energy\cite{e-ASTROGAM:2016bph}. At energies below (above) 10 MeV, this instrument relies primarily on Compton scattering (pair conversion).}
\label{astrogam}
\end{figure}

To produce a simulated data set, we first convolve each of the templates by the point spread function of e-ASTROGAM, which we treat as a gaussian with a 68\% containment radius as given in the upper frame of Fig.~\ref{astrogam}~\cite{e-ASTROGAM:2016bph}. Then, after summing the templates, we calculate the mean number of events in a given angular and energy bin by multiplying the flux in that bin by the acceptance of e-ASTROGAM (as given in the lower frame of Fig.~\ref{astrogam})~\cite{e-ASTROGAM:2016bph}, and by five years of observation time. For each bin, we then randomly draw from a Poisson distribution with the appropriate mean number of events to find the simulated number of events in that bin. Once we have a simulated data set for a given choice of $m_{\rm BH}$, $f_{\rm BH}$, and $\gamma$, we can calculate the likelihood for a model described by any given sum of the templates listed above.

\section{Projected Constraints}

To derive the projected constraints for e-ASTROGAM (or a similar instrument) on the abundance of primordial black holes, we simulate a data set assuming that no such black holes are present. Then, for each choice of $m_{\rm BH}$ and $\gamma$, we calculate the likelihood as a function of $f_{\rm BH}$, in order to place an upper limit on $f_{\rm BH}$. To identify the points in parameter space with the maximum likelihood, and to derive the appropriate confidence intervals around those points, we utilize the publicly available MINUIT algorithm~\cite{James:1975dr}. Because MINUIT can occasionally identify false-minima, we use the PyMultiNest package~\cite{Buchner:2014nha} to test the robustness of our results by searching for global minima which may not have been encountered in our MINUIT scan.


\begin{figure*}
\includegraphics[width=3.25in,angle=0]{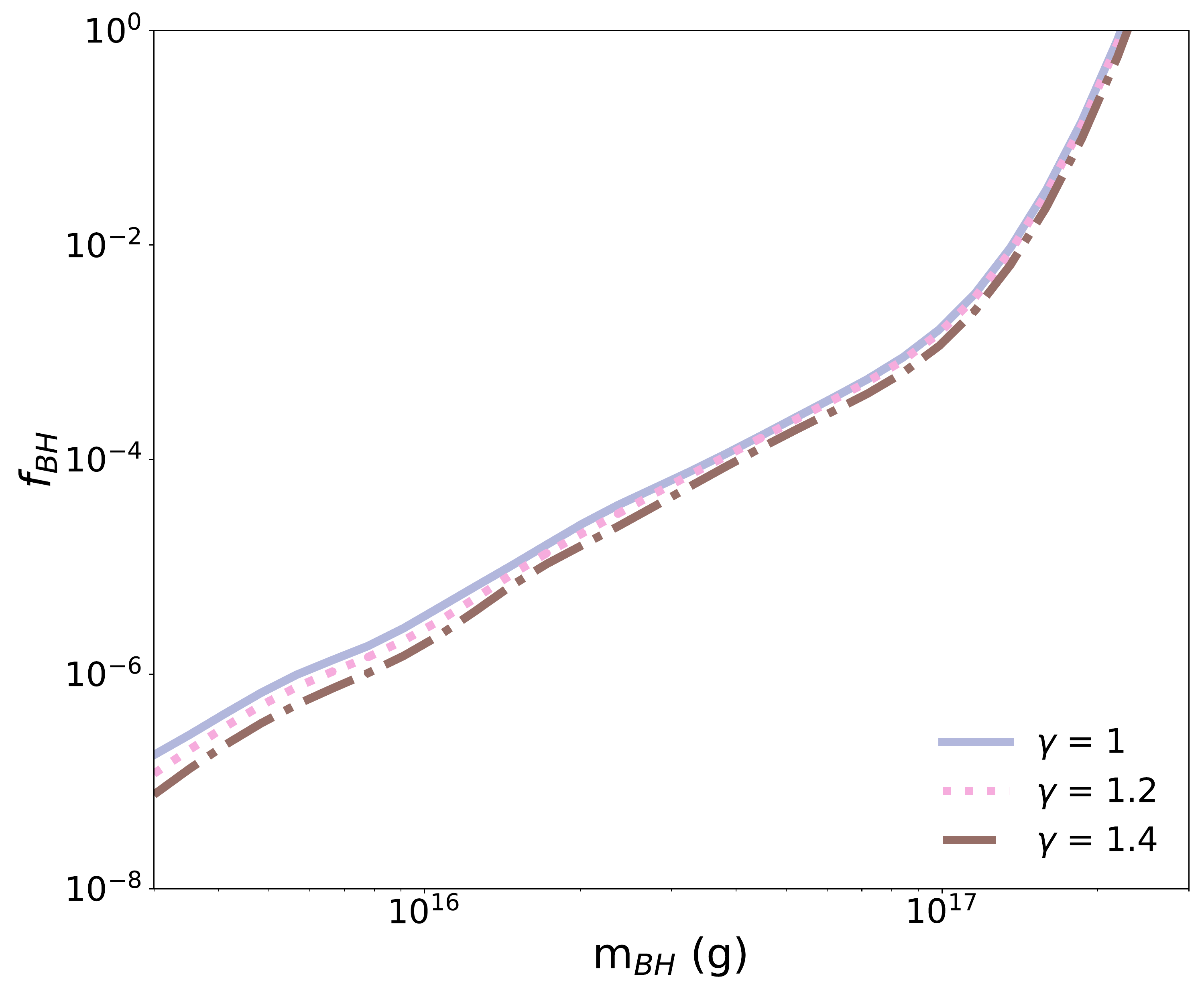}
\includegraphics[width=3.25in,angle=0]{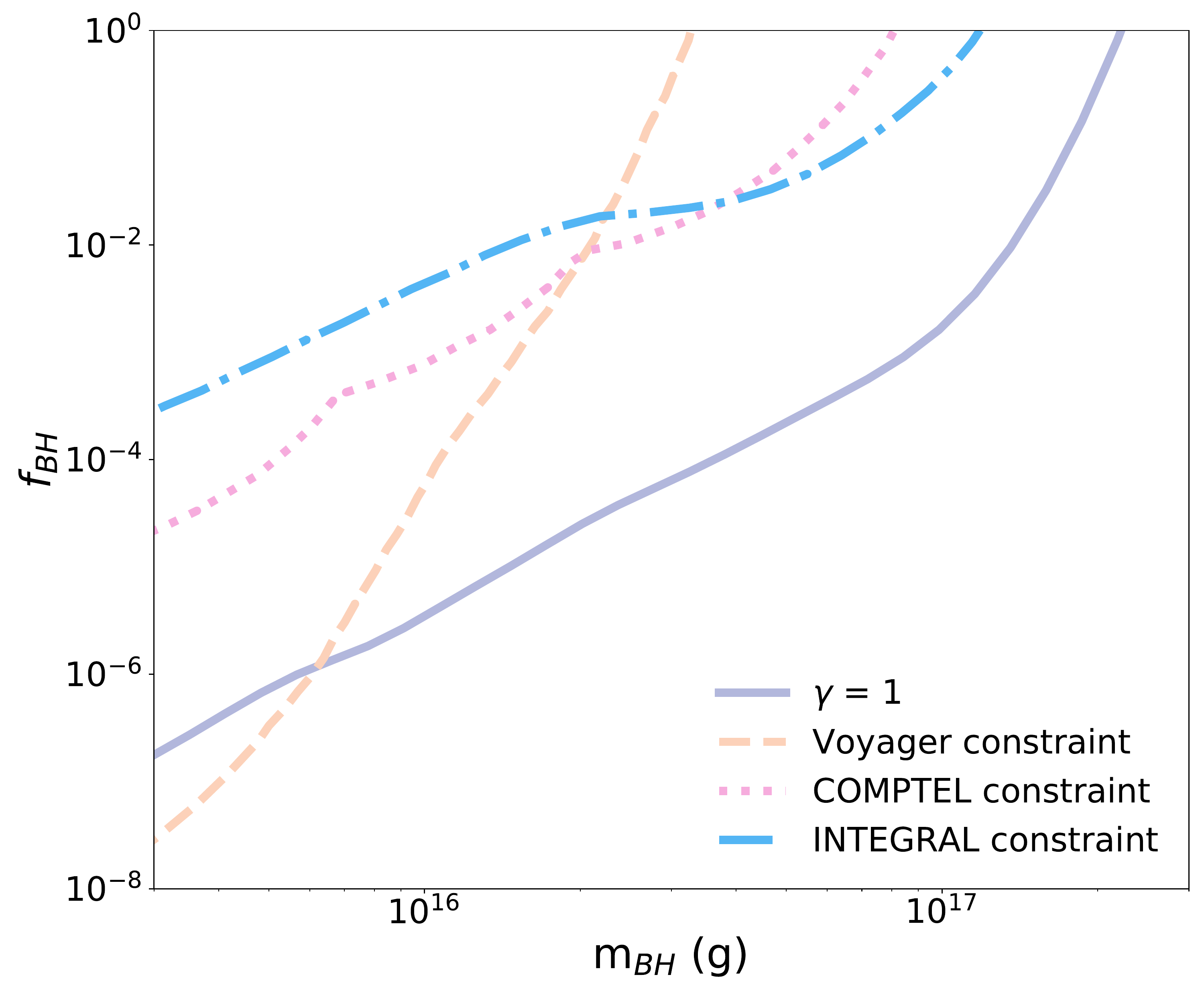}
\caption{Our projected 95\% confidence level upper limits on the fraction of the dark matter that could consist of primordial black holes, $f_{\rm BH}$, after 5 years of observation with e-ASTROGAM. In the left frame, we show results for black holes that are distributed according to a generalized NFW profile with $\gamma=1.0$, 1.2, or 1.4. In the right frame, our projected constraints are compared to existing constraints derived from local measurements of the cosmic-ray electron-positron flux by the Voyager 1 satellite~\cite{Boudaud:2018hqb}, and gamma-ray observations of the Inner Galaxy by COMPTEL and INTEGRAL~\cite{Keith:2021guq}.}
\label{constraintNFW}
\end{figure*}

\begin{figure}
\includegraphics[width=3.25in,angle=0]{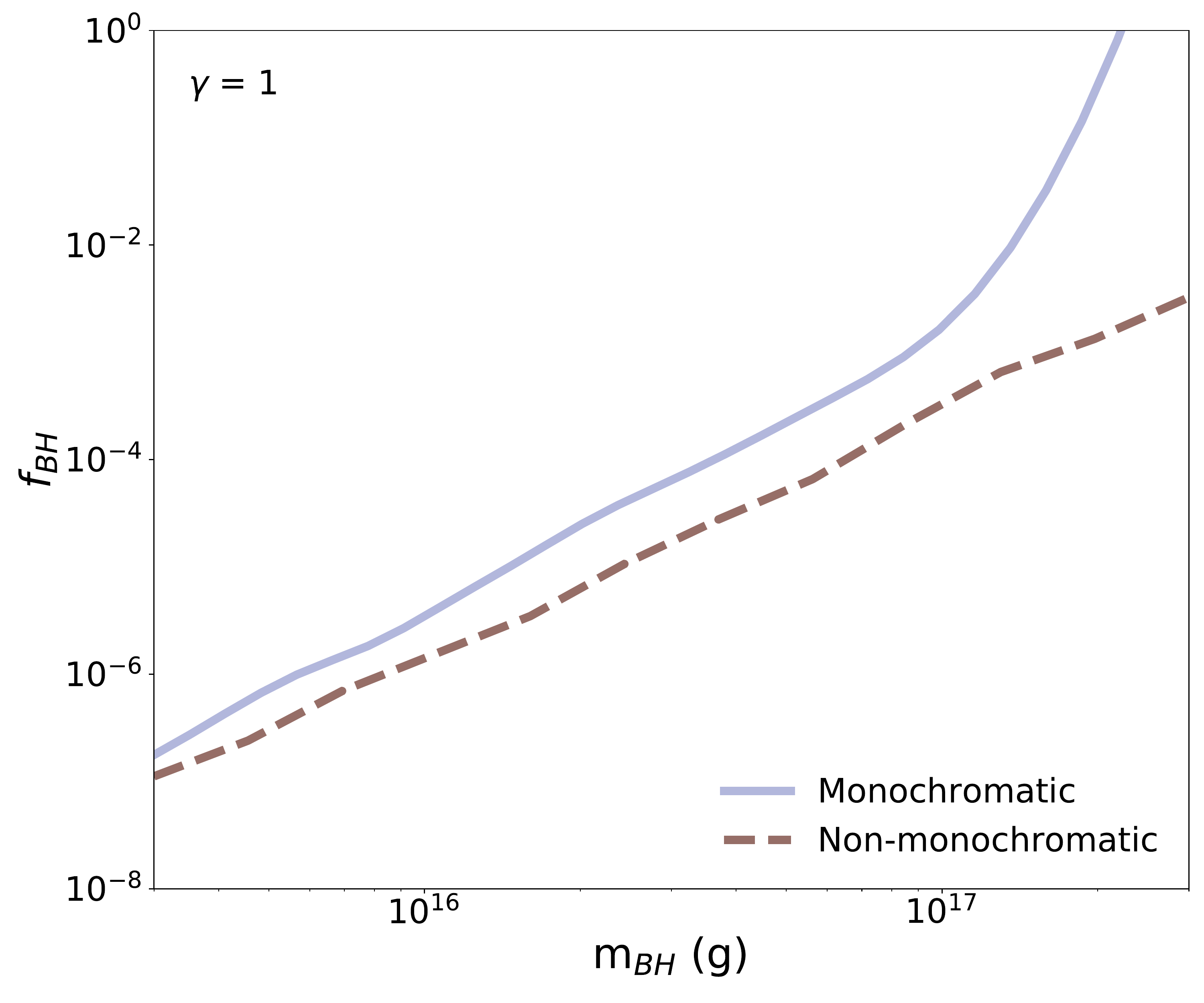}
\caption{The solid curve in this figure represents the same constraint as that shown in Fig.~\ref{constraintNFW} (for the case of $\gamma=1$), while the dashed curve is that obtained for a population of black holes with masses that are distributed according to a log-normal distribution with a variance of $\sigma=2$ and that is centered around $\mu = \ln (m_{\rm BH})$.}
\label{lognormal}
\end{figure}

In the left frame of Fig.~\ref{constraintNFW}, we show our 95\% confidence-level projected upper limits on $f_{\rm BH}$ for black holes distributed according to a generalized NFW profile with $\gamma=1.0$, 1.2, or 1.4.\footnote{To reduce the impact of stochastic variations in our simulated data sets, we show as our projected constraints the average result obtained over five independent realizations.} In the right frame of this figure, we compare this constraint to those previously derived from Voyager 1~\cite{Boudaud:2018hqb}, as well as COMPTEL and INTEGRAL~\cite{Keith:2021guq}, for the specific case of $\gamma=1$. For black holes more massive than $m_{\rm BH} \sim 6 \times 10^{15} \, {\rm g}$, our projected constraints would represent the most stringent limits on the Hawking radiation from black holes. 

Up to this point in our analysis, we have adopted a monochromatic distribution for the masses of the black holes. More realistically, we might expect a population of primordial black holes to contain members with a range of different masses. To this end, we repeated our calculation considering black holes that are distributed according to a log-normal distribution with a variance of $\sigma=2$. These results, which are shown in Fig.~\ref{lognormal}, are somewhat more stringent those obtained for the case of a monochromatic mass distribution.

Ideally, one would independently float the intensity and spectrum from each gamma-ray point source in a template-based analysis. Computational limitations, however, make such an approach unrealistic. For this reason, we have adopted in our calculations a single template to account for all of the known gamma-ray point sources in this region of the sky. Most of these sources are very morphologically distinct from our black hole template, making it very unlikely that this choice would significantly impact our projections. One might speculate, however, that an individual point source located near the Galactic Center could be partially degenerate with our black hole template, potentially biasing our results. To test this possibility, we have repeated our analysis including an additional template to account for the relatively bright and centrally-located point source 4FGL J1745.6-2859, which is associated with the Milky Way's supermassive black hole, Sgr A$^*$. The constraints obtained in this way differ negligibly from those shown in Fig.~\ref{constraintNFW}, never by more than a few percent, thus indicating that the emission from individual point sources is unlikely to be confused with that from primordial black holes in our analysis.

\section{Sensitivity to PBHs Capable of Generating the 511 keV Excess}

Measurements of the Inner Milky Way by the INTEGRAL satellite have identified an excess of 511 keV photons, consisting of a flux of (1.07 $\pm$ 0.03) $\times$10$^{-3}$ photons cm$^{-2}$ s$^{-1}$ and corresponding to the injection of $\sim$\,$2 \times 10^{43}$ positrons per second~\cite{Weidenspointner:2004my, Churazov:2004as, Weidenspointner:2007rs, Jean:2005af, Weidenspointner:2008zz, Kierans:2019aqz, Prantzos:2005pz}. While various astrophysical sources of this emission have been considered~\cite{Kalemci:2006bz,Casse:2003fh,Bertone:2004ek,Guessoum:2006fs,Bartels:2018eyb,Takhistov:2019zyb,Fuller:2018ttb}, these interpretations each face considerable challenges (for a review, see Ref.~\cite{Prantzos:2010wi}). In light of this situation, a number of more exotic scenarios have been proposed, including those in which the 511 keV excess is produced by the annihilation~\cite{Boehm:2003bt,Huh:2007zw,Hooper:2008im,Khalil:2008kp},  decay~\cite{Hooper:2004qf,Cembranos:2008bw,Craig:2009zv}, or upscattering~\cite{Finkbeiner:2007kk,Pospelov:2007xh,Cline:2010kv,Cline:2012yx} of dark matter particles, or by Q-balls~\cite{Kasuya:2005ay}, pico-charged particles~\cite{Farzan:2017hol,Farzan:2020llg}, quark nuggets~\cite{Lawson:2016mpu}, or 
unstable MeV-scale states produced in supernovae~\cite{Davoudiasl:2009ud}. It is also possible that the excess of 511 keV photons could be produced through the Hawking evaporation of a population of primordial black holes concentrated in the Inner Galaxy~\cite{Frampton:2005fk,Bambi:2008kx,Cai:2020fnq,Keith:2021guq} (see also, Refs.~\cite{Laha:2019ssq,DeRocco:2019fjq}). In particular, a population of black holes with a distribution of masses that peaks around $m_{\rm BH} \sim (1-4) \times 10^{16}$ g could plausibly generate this signal if they are distributed in a very concentrated profile around the Galactic Center~\cite{Keith:2021guq}.

The observed morphology of the 511 keV excess~\cite{Bouchet:2010dj} is quite steeply concentrated around the Galactic Center. As a result, if primordial black holes are to generate these excess photons, they must be distributed with a profile that is at least as centrally concentrated as $\gamma \sim 1.6$~\cite{Keith:2021guq} (see also, Refs.~\cite{Vincent:2012an,Ascasibar:2005rw}). This is significantly steeper than the profiles favored by numerical simulations of cold dark matter, which typically favor $\gamma \sim 1.0-1.4$~\cite{Gnedin:2004cx, Governato_2012, Kuhlen:2012qw, Weinberg:2001gm, Weinberg:2006ps, Calore:2015oya, Schaller:2014uwa, DiCintio:2014xia, DiCintio:2013qxa, Schaller:2015mua, Bernal:2016guq}. Such a scenario thus requires a greater degree of adiabatic contraction than is suggested by current simulations (see, for example, Ref.~\cite{Gnedin:2011uj}).

\begin{figure}
\includegraphics[width=3.25in,angle=0]{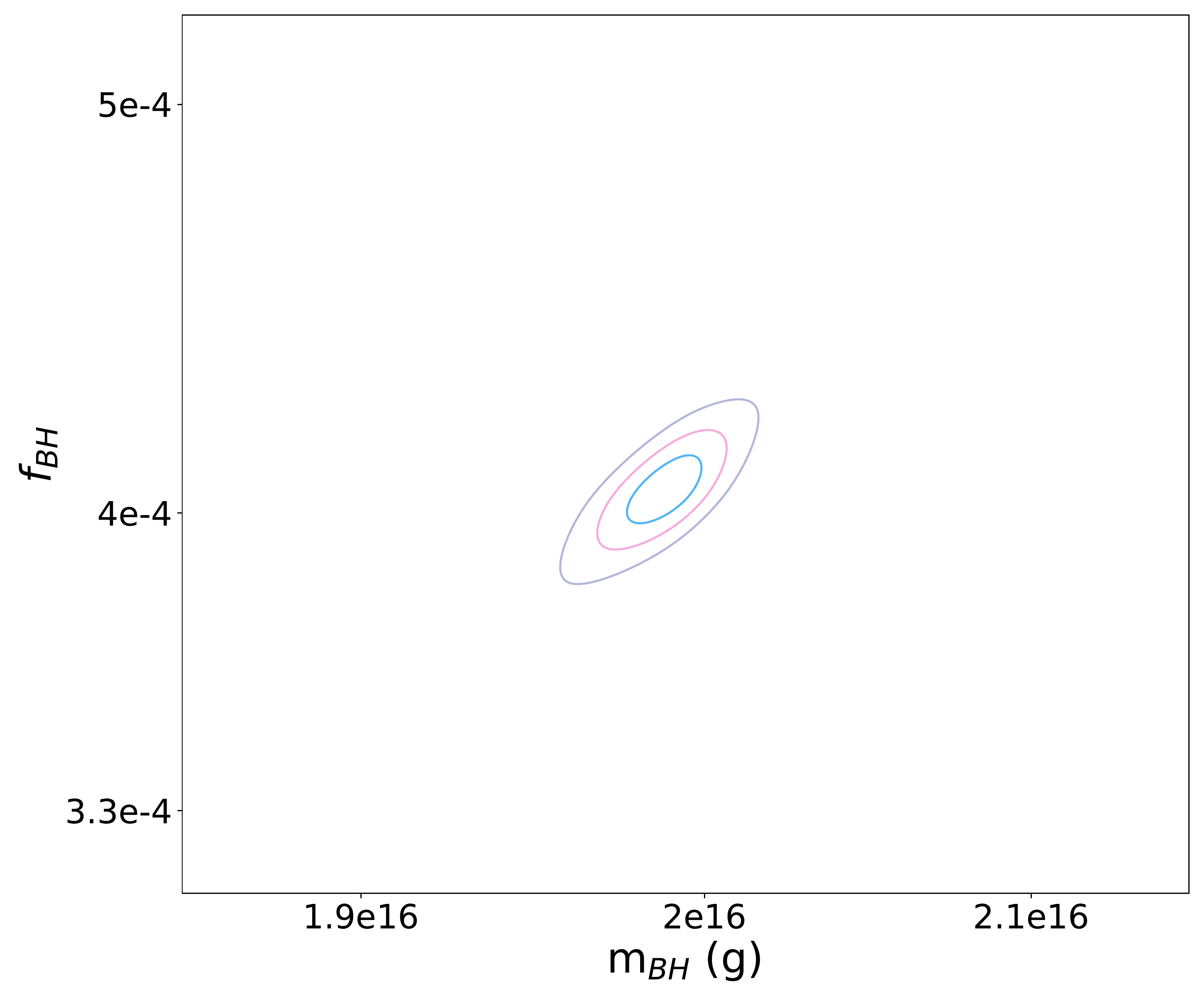}
\caption{The ability of e-ASTROGAM to measure the properties of a black hole population in a scenario in which $m_{\rm BH}=2 \times 10^{16} \, {\rm g}$, $f_{\rm BH} = 4 \times 10^{-4}$, and $\gamma=1.6$, as motivated by the 511 keV excess observed by INTEGRAL~\cite{Keith:2021guq}. The contours reflect the projected 1$\sigma$, 2$\sigma$, and 3$\sigma$ constraints of these quantities.}
\label{contours}
\end{figure}

To project the sensitivity of e-ASTROGAM to a population of black holes that could be responsible for the observed 511 keV excess, we simulate a data set for the case of $m_{\rm BH}=2 \times 10^{16} \, {\rm g}$, $f_{\rm BH} = 4 \times 10^{-4}$, and $\gamma=1.6$~\cite{Keith:2021guq}. We then calculate the maximum value of the likelihood that is obtained as a function of these three parameters. In Fig.~\ref{contours}, we show the results of this analysis. This figure demonstrates that an instrument such as e-ASTROGAM would not only be able to detect the Hawking radiation from a black hole population responsible for the 511 keV excess, but would be able to characterize the properties of such a population with remarkable precision. While such a result could be impacted by systematic uncertainties that we have not accounted for in our analysis, we consider it clear that e-ASTROGAM would be able to quite accurately detect and measure the gamma-ray emission produced by such a population of primordial black holes.

\section{Summary and Conclusion}
  
In this study, we have evaluated the ability of future MeV-scale gamma-ray telescopes such as e-ASTROGAM or AMEGO to detect and characterize the Hawking radiation from a population of primordial black holes located in the inner volume of the Milky Way. To this end, we have calculated the gamma-ray emission from black holes, including contributions from direct Hawking radiation, inflight positron annihilation, and final state radiation. We then performed an analysis utilizing a series of spatial templates, allowing us to fully exploit the morphological and spectral information provided by such an instrument. We have included in our analysis templates associated with pion production, inverse Compton scattering, bremsstrahlung, known point sources, the Galactic Center gamma-ray excess, and the extragalactic gamma-ray background, as well as that associated with the Hawking radiation from a population of primordial black holes.  

At the present time, the strongest constraints on Hawking radiation come from the Voyager 1, COMPTEL, and INTEGRAL satellites~\cite{Keith:2021guq,Coogan:2020tuf,Laha:2020ivk,Boudaud:2018hqb,Dasgupta:2019cae}. More specifically, local measurements of the cosmic-ray electron-positron flux by Voyager 1 provide the strongest constraint on black holes lighter than $m_{\rm BH }\sim (1-2) \times 10^{16} \, {\rm g}$~\cite{Boudaud:2018hqb}, while MeV-scale gamma-ray observations of the Inner Galaxy by COMPTEL and INTEGRAL provide the leading constraints in the mass range of $m_{\rm BH}\sim 10^{16}-10^{17} \, {\rm g}$~\cite{Keith:2021guq,Coogan:2020tuf,Laha:2020ivk,Coogan:2020tuf}. In the absence of a black hole population, we project that e-ASTROGAM will be able to provide the strongest constraints on black holes in the mass range of \mbox{$m_{\rm BH} \sim (0.6-20) \times 10^{16} \, {\rm g}$}. Over much of this mass range, the sensitivity of e-ASTROGAM will exceed that of existing or past experiments by roughly two orders of magnitude. 

It has been previously pointed out that primordial black holes could be responsible for the excess of 511 keV photons observed from the Inner Galaxy by the INTEGRAL satellite. This requires the mass distribution of the black hole population to peak at around $m_{\rm BH} \sim (1-4) \times 10^{16} \, {\rm g}$, and for these objects to be distributed in a very concentrated profile around the Galactic Center~\cite{Keith:2021guq}. In such a scenario, we find that an instrument such as AMEGO or e-ASTROGAM would not only clearly detect the Hawking radiation from such a population, but would be able to quite precisely measure the abundance and mass distribution of the responsible black holes.

While the results presented here were calculated using the acceptance and angular resolution of the proposed e-ASTROGAM experiment, similar results could be obtained for other designs with comparable sensitivity to MeV-scale gamma rays, such as the proposed satellite-based mission, AMEGO~\cite{McEnery:2019tcm}.

\begin{acknowledgments}  

CK would like to thank Cory Cotter for helpful discussions. CK is supported by the National Science Foundation Graduate Research Fellowship Program under Grant No. DGE-1746045. DH is supported by the Fermi Research Alliance, LLC under Contract No.~DE-AC02-07CH11359 with the U.S. Department of Energy, Office of Science, Office of High Energy Physics. TL is partially supported by the Swedish Research Council under contract 2019-05135, the Swedish National Space Agency under contract 117/19 and the European Research Council under grant 742104.

\end{acknowledgments}

\bibliography{eastrogam2022pref}

\end{document}